\documentclass[]{AO4ELT}  

\usepackage[flushleft]{threeparttable}
\usepackage{microtype}
\usepackage{biblatex}
\usepackage{amsmath,amsfonts,amssymb}
\usepackage{graphicx}
\usepackage{pst-all} 
\usepackage[colorlinks=true, allcolors=blue]{hyperref}
\usepackage{amsmath}
\usepackage{graphicx}
\usepackage[caption=false]{subfig}
\usepackage{float}
\usepackage{multirow,array,booktabs}

\DeclareUnicodeCharacter{2212}{-}
\usepackage{tabularx}
\usepackage{amsmath,amsfonts,amssymb}
\usepackage{graphicx}
\usepackage[colorlinks=true, allcolors=blue]{hyperref}

\makeatletter         
\def\@maketitle{
\includegraphics[width = 170mm]{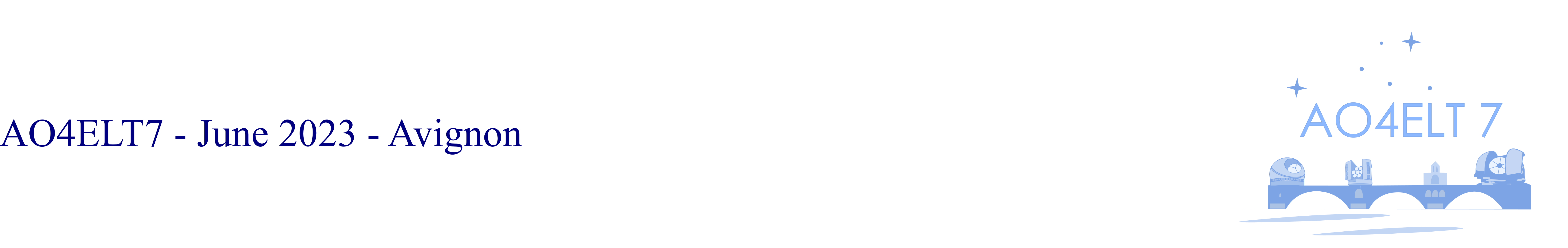}\\[8ex]
\begin{center}
{\Huge \bfseries \sffamily \@title }\\[4ex] 
{\Large  \@author}\\[4ex] 
\@date
\end{center}}

\title{GPI 2.0: Performance Evaluation of the Wavefront Sensor's  EMCCD}

\author[a]{Clarissa R. Do Ó}
\author[a]{Saavidra Perera}
\author[a]{Jérôme Maire}
\author[a]{Jayke S. Nguyen}
\author[j]{Vincent Chambouleyron}
\author[a]{Quinn M. Konopacky}
\author[b]{Jeffrey Chilcote}
\author[c]{Joeleff Fitzsimmons}
\author[b]{Randall Hamper}
\author[c]{Dan Kerley}
\author[j]{Bruce Macintosh}
\author[c]{Christian Marois}
\author[e]{Fredrik Rantakyrö}
\author[f]{Dmitry Savranksy}
\author[c]{Jean-Pierre Veran}
\author[g]{Guido Agapito}
\author[h]{S. Mark Ammons}
\author[g]{Marco Bonaglia}
\author[i]{Marc-Andre Boucher}
\author[c]{Jennifer Dunn}
\author[g]{Simone Esposito}
\author[i]{Guillaume Filion}
\author[i]{Jean Thomas Landry} 
\author[c]{Olivier Lardiere}
\author[f]{Duan Li}
\author[d]{Alex Madurowicz}
\author[b]{Dillon Peng}
\author[h]{Lisa Poyneer}
\author[b]{Eckhart Spalding}

\affil[a]{Center for Astrophysics and Space Science, University of California San Diego, La Jolla, CA
92093, USA}
\affil[b]{Department of Physics, University of Notre Dame, 225 Nieuwland Science Hall, Notre Dame, IN 46556, USA}
\affil[c]{National Research Council of Canada Herzberg, 5071 West Saanich Rd, Victoria, BC, V9E 2E7, Canada}
\affil[d]{Kavli Institute for Particle Astrophysics and Cosmology, Stanford University, Stanford, CA 94305, USA}
\affil[e]{Gemini Observatory, 670 N. A’ohoku Place, Hilo, HI 96720, USA}
\affil[f]{Sibley School of Mechanical and Aerospace Engineering, Cornell University, Ithaca, NY 14853, USA}
\affil[g]{Arcetri, Largo Enrico Fermi 5, I - 50125 Florence, Italy}
\affil[h]{Lawrence Livermore National Laboratory, Livermore, CA 94551, USA}
\affil[i]{Opto-Mécanique de Précision, 146 Bigaouette St. Quebec City, QC, Canada, G1K 4L2}
\affil[j]{Center for Adaptive Optics, University of California Santa Cruz, Santa Cruz, CA 95064, USA}

\authorinfo{Further author information: (Send correspondence to C.D.O) \\ E-mail:cdoo@ucsd.edu}

\begin{document} 
\maketitle

\begin{abstract}
The Gemini Planet Imager (GPI) is a high contrast imaging instrument that aims to detect and characterize extrasolar planets. GPI is being upgraded to GPI 2.0, with several subsystems receiving a re-design to improve the instrument's contrast. To enable observations on fainter targets and increase stability on brighter ones, one of the upgrades is to the adaptive optics system.  The current Shack-Hartmann wavefront sensor (WFS) is being replaced by a pyramid WFS with an low-noise electron multiplying CCD (EMCCD). EMCCDs are detectors capable of counting single photon events at high speed and high sensitivity. In this work, we characterize the performance of the HNü 240 EMCCD from Nüvü Cameras, which was custom-built for GPI 2.0. The HNü 240 EMCCD's characteristics make it well suited for extreme AO: it has low dark current ($<$ 0.01 e-/pix/fr), low readout noise (0.1 e-/pix/fr at a gain of 5000), high quantum efficiency (\>90\% at wavelengths from 600-800 nm; \>70\% from 800-900 nm), and fast readout (up to 3000 fps full frame). Here we present test results on the EMCCD’s noise contributors, such as the readout noise, pixel-to-pixel variability and CCD bias. We also tested the linearity and EM gain calibration of the detector. All camera tests were conducted before its integration into the GPI 2.0 PWFS system.
\end{abstract}


\section{INTRODUCTION}
Searching for new directly imaged planetary systems can help shed light on exoplanet formation and orbital architecture. One key aspect of finding new exoplanets is the development of state-of-the-art high contrast imaging instruments.
Directly imaged planets are much fainter than their host stars, which is why high contrast imaging instruments require a coronagraph, a device that blocks the starlight such that an off-axis signal, like the planet's light, can be detected \cite{Milli2017}. However, coronagraphs require the starlight to be directly aligned with the device, such that wavefront aberrations (in particular tip or tilt) can severely suppress the contrast capabilities of an instrument. The contrast needed for current instruments to detect Jupiter-mass planets is large, which requires an efficient wavefront measurement and correction using a highly sensitive and efficient wavefront sensor. \par
The Gemini Planet Imager (GPI) is a high contrast imaging instrument. It operated at Gemini North for 6 years, with the main goal of finding and characterizing Jupiter mass exoplanets in wide orbits. GPI was designed to perform a statistical analysis of a number of wide orbit gas giants, in the hopes of constraining the formation mechanisms of the planetary systems they inhabit. \par
\begin{figure} [ht]
   \begin{center}
   \begin{tabular}{c} 
   \includegraphics[height = 7cm]{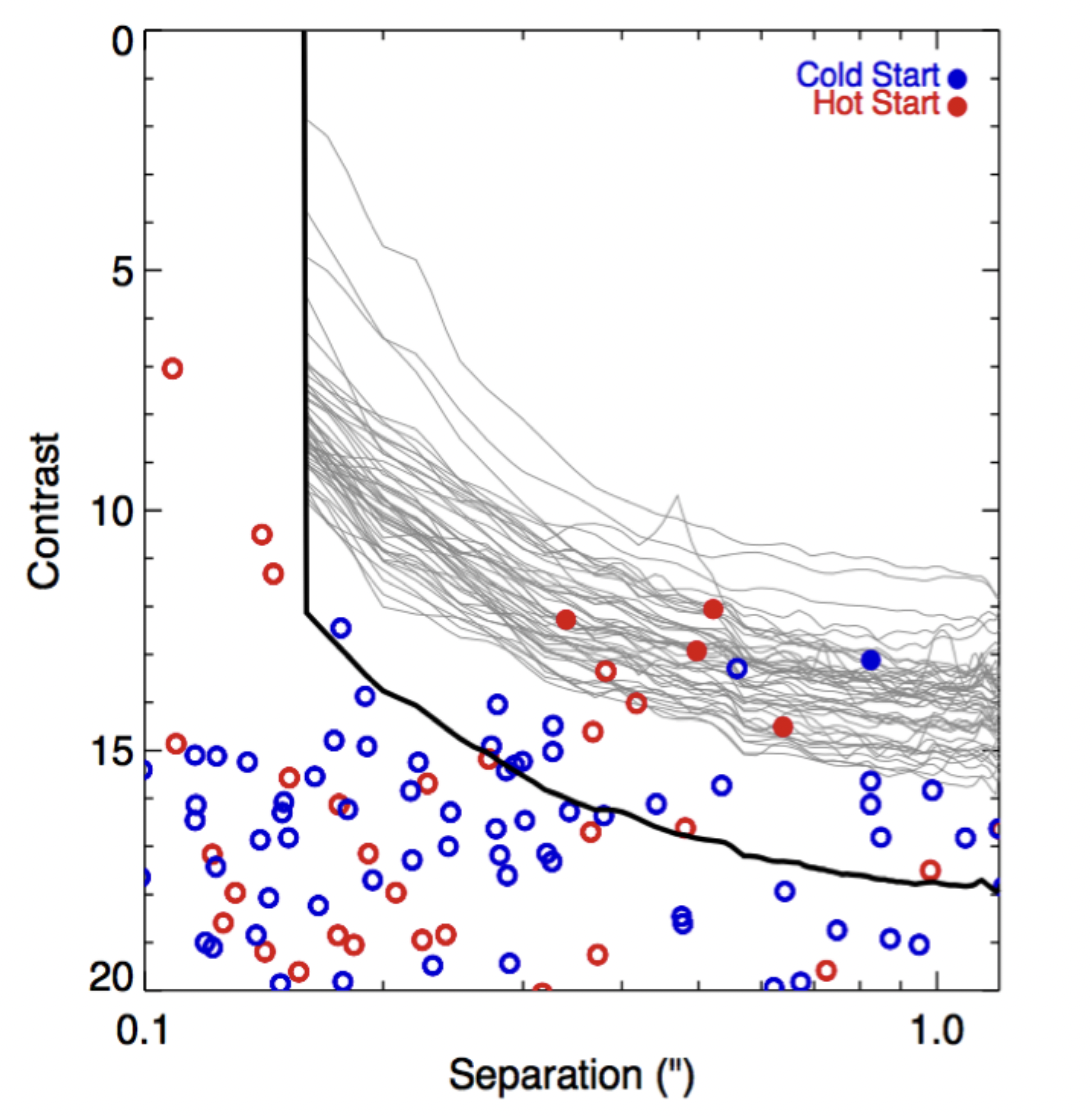}
   \end{tabular}
   \end{center}
   \caption[example] 
   {\label{fig:chilcote.png} 
GPI's contrast curve (in magnitudes) as a function of separation in arcseconds from the host start. Typical GPI 1.0 contrast curves are presented in light gray, with the GPI 2.0 predicted contrast curve represented in black. The GPI 2.0 will reach planets 3.2 magnitudes fainter than GPI 1.0, allowing for many more ``cold-start" and closer-in planets to be detected. The filled in circles are planets that could be found by GPI 1.0's current set-up, while hollow circles represent planets that would fall below the current contrast curve for the host star. The planets come from a simulated exoplanet population. Figure from \cite{Chilcote2018}.}
   \end{figure}

The improvement in high contrast imaging technologies can allow for better contrasts at smaller separations from the host star, enabling the detection of closer-in planets and planets more consistent with the core accretion model  \cite{Chilcote2018}, \cite{Nielsen_2019}. The core accretion model is also named cold-start because the planet's initial entropy is much lower (thus making it less luminous) than that of a hot-start (or gravitational instability) formation, where the rapid formation of clumps causes the planet to possess a higher entropy and consequently a higher luminosity \cite{Marley2007}. 
GPI is going through an upgrade to become GPI 2.0. GPI 2.0 aims to have higher contrast than GPI 1.0, giving it access to companions that are fainter and closer to their host stars as well as fainter stars, which allows for an increased amount of young targets not detectable with GPI 1.0. In order to achieve this goal, several subsystems are receiving upgrades, including the calibration unit, the coronagraphic system, the integral field spectrograph (IFS) and the adaptive optics system (AO). The predicted contrast after the upgrade to GPI 2.0 is shown in Figure \ref{fig:chilcote.png}. It can be seen that GPI 2.0 will unlock the detection of planets consistent with the cold-start models (GPI 1.0 was already quite sensitive to hot-start planets \cite{Chilcote2018}), enabling detections at the peak of the giant planet population distribution (\cite{Nielsen_2019}). \par
This work is organized as follows: in section \ref{pwfs} the GPI 2.0 wavefront sensor design is introduced. In section \ref{emccd}, the wavefront sensor's EMCCD is introduced. Our test results are shown in section \ref{tests}, with readout noise results in section \ref{ron}, clock-induced charges in section \ref{cic}, multiple regions of interest in section \ref{mroi}, binning in section \ref{binning} and full frame rate results in section \ref{ffr}. The flat test results are presented in section \ref{flats}, divided in EM gain linearity (section \ref{emgainlinear}) and exposure time linearity (sections \ref{timelinearity} and \ref{ratios}) tests. We discuss and conclude our tests in section \ref{conclusion}.

\subsection{The Gemini Planet Imager 2.0: Wavefront Sensor Upgrade} \label{pwfs}

\begin{figure} [ht]
   \begin{center}
   \begin{tabular}{c} 
   \includegraphics[width = 12cm]{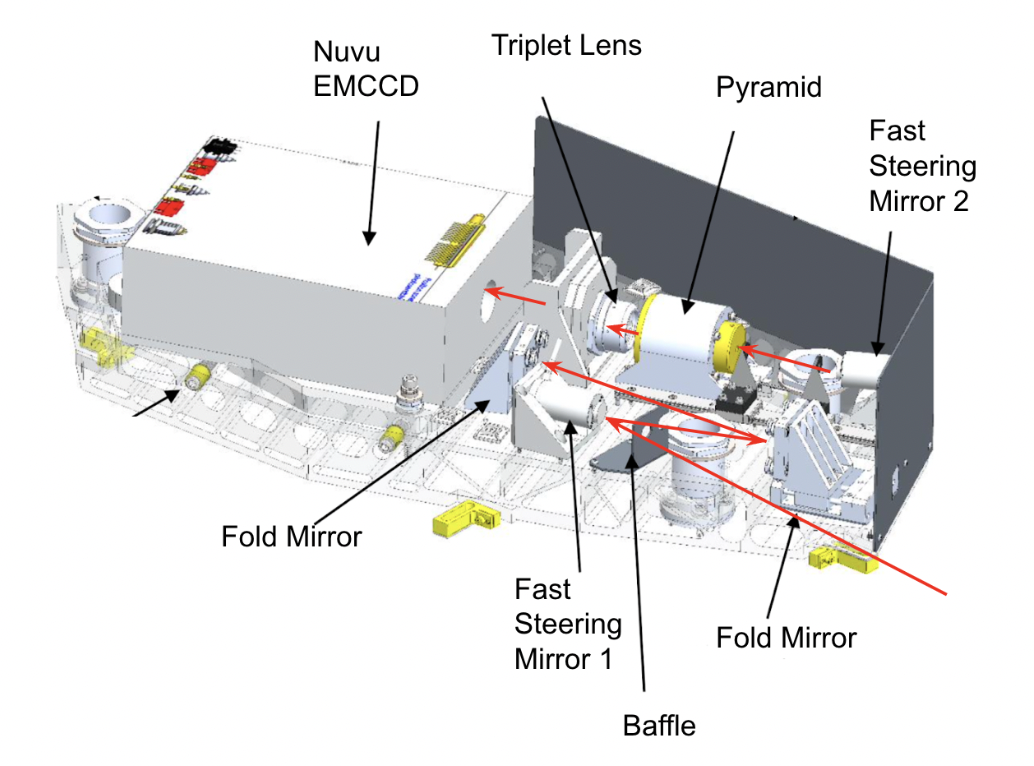}
   \end{tabular}
   \end{center}
   \caption[example] 
   {\label{fig:pwfs} 
GPI 2.0's pyramid WFS was designed by the Herzberg Astronomy and Astrophysics Research Center (HAA) and simulated by Stanford University, with the assembly taking place at UCSD. The design includes two fast steering mirrors, one for modulation (FSM 1) and one for tip/tilt adjustments (FSM 2). Between the two FSMs, there are two fold mirrors; the first fold mirror also acts as a focus stage. After FSM 2, the light goes through the double four-sided pyramid and through a camera lens before reaching the Nüvü EMCCD.}
   \end{figure}

At the University of California, San Diego (UCSD), the AO subsystem is being upgraded. Specifically, GPI 2.0 will replace the current Shack-Hartmann WFS to a pyramid WFS, because its higher sensitivity to low-order aberrations will allow for the detection of fainter targets. 
GPI 2.0's wavefront sensor design is shown in Figure \ref{fig:pwfs}. It contains a fast steering mirror which modulates the light that hits the pyramid to increase the dynamic range of the WFS. It also has a fold mirror which allows for focusing (focus stage) and a second fast steering mirror for tip/tilt adjustments. The light goes through the double four-sided pyramid and a camera lens before hitting the detector, an EMCCD.

\subsection{The EMCCD} \label{emccd}

\begin{figure} [ht]
   \begin{center}
   \begin{tabular}{c} 
   \includegraphics[width = \columnwidth]{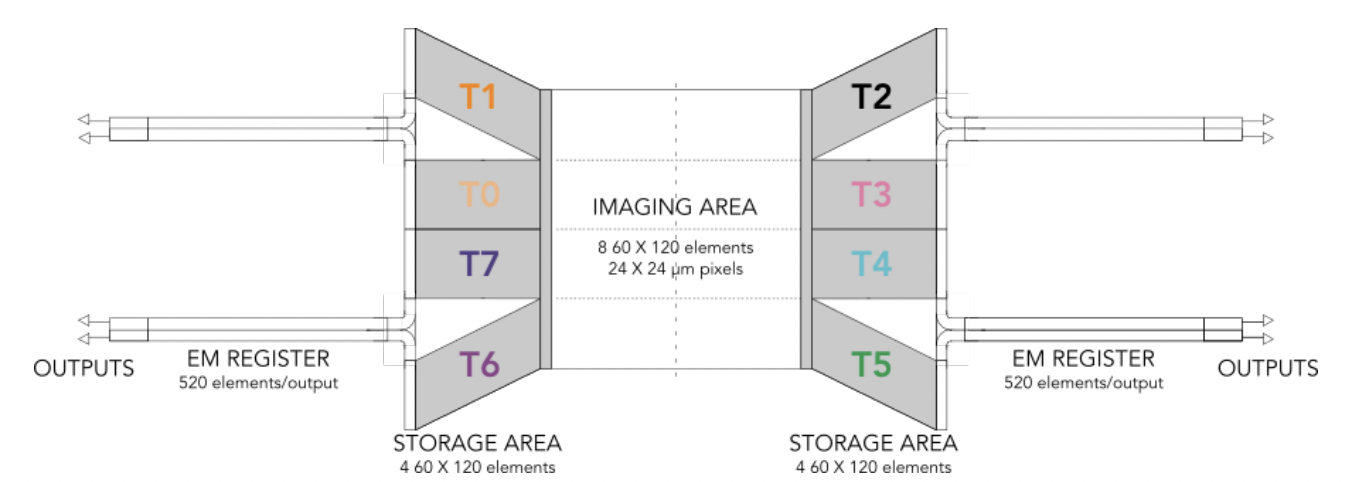}
   \end{tabular}
   \end{center}
   \caption[example] 
   {\label{fig:emccdchip} 
The EMCCD's chip. The imaging area is composed of 8 60x120 imaging areas (``outputs"), forming a 240x240 pixel image, storage areas and multiplication (``EM") registers. The camera has a nominal temperature operation of -45 $^{\circ}$C. Figure is from \cite{10.1117/12.2626240}.}. 
   \end{figure}

The WFS's detector is an electron-multiplying CCD (EMCCD) manufactured by Nüvü Cameras \cite{Nuvu_Proceedings}. EMCCDs are detectors capable of counting single photon events at high speed and high sensitivity. The EMCCD chip configuration is presented in Figure \ref{fig:emccdchip}. Much like a traditional CCD, the EMCCD turns photons into electrons via the photoelectric effect. However, unlike the traditional CCD, the EMCCD presents an extra register called the multiplication register. Once the photons hit the silicon body of the chip, the electrons in the imaging area travel row by row to the storage area. This mechanism allows for the next frame to be taken while the previous one is being processed \cite{NuVuCameras}. Once in the storage section, the electrons travel to the multiplication register where hundreds of electrodes accelerate them, causing a phenomenon called impact ionization. Using high voltages, the captured electrons collide with the multiplication registers' silicon atoms, ripping an electron from the atom. This new electron then becomes part of the measured signal \cite{NuVuCameras}.\par
\begin{figure} [ht]
   \begin{center}
   \begin{tabular}{c} 
   \includegraphics[height=5cm]{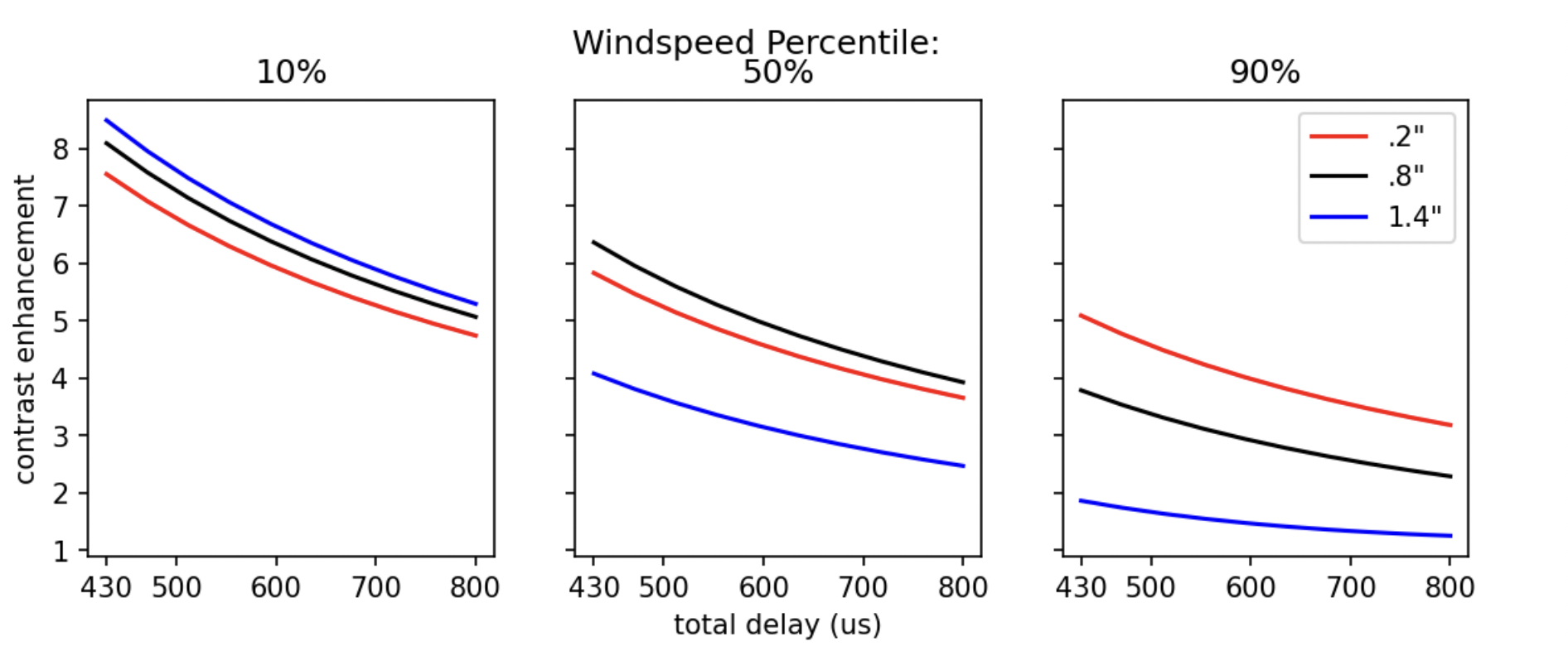}
   \end{tabular}
   \end{center}
   \caption[example] 
   {\label{fig:alexplot} 
Simulated contrast enhancement as a function of total delay (readout time of AO camera and real time control) for GPI 2.0. This plot was made using an end-to-end simulation of contrast curves for GPI 2.0's design. Figure from \cite{Madurowicz2020}.} 
   \end{figure}

The EM gain sets how much an electron signal will be multiplied by, which is achieved by changing the voltage in the multiplication register. This specific EMCCD chip, Teledyne e2v CCD220, has 8 outputs for a faster readout of these electrons, with 2 outputs sharing one multiplication register. EMCCDs are particularly useful for AO systems because of their high operating speed of 3,000 FPS (thus measuring the fast changing atmosphere) and their high sensitivity and low noise, which allows for better corrections of the wavefront to high order aberrations (\cite{10.1117/12.2626240}, \cite{doi:10.1142/S2251171719500016}). Figure \ref{fig:alexplot} shows how the contrast enhancement relies on the total delay, defined as the WFS camera's readout time and the real-time control. For any windspeed percentile, contrast is lost if the camera's readout is not fast enough to keep up with the atmosphere. \par
The EMCCD for GPI 2.0 has two CameraLink cables that connect to a Pleora iPORT CL-Ten External Frame Grabber. The Pleora is connected to our laboratory computer using 10 Gb/s ethernet cable.

\section{Conducted Tests} \label{tests}
\begin{figure} [ht]
   \begin{center}
   \begin{tabular}{c} 
   \includegraphics[height=7cm]{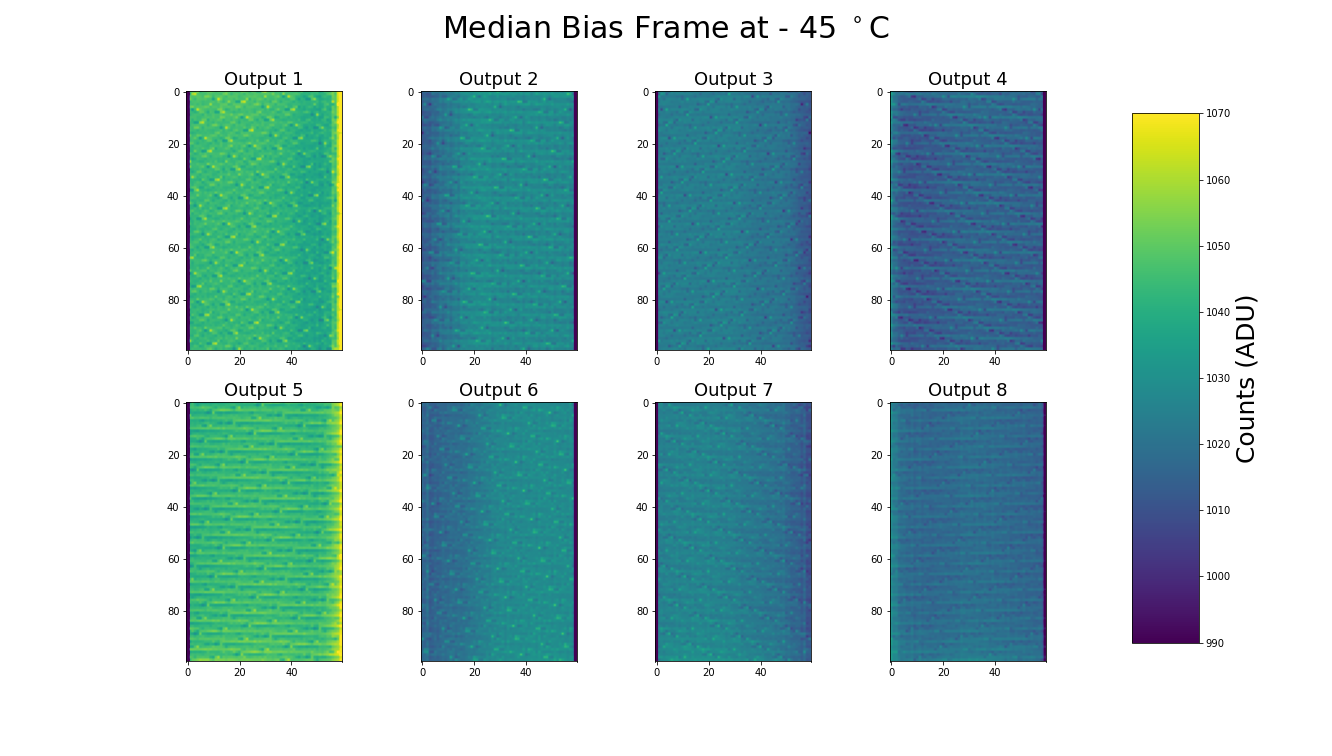}
   \end{tabular}
   \end{center}
   \caption[example] 
   {\label{fig:outputrep} 
Representation of the 8 outputs of our camera. We number them from 1-8 as shown in the Figure. In order to better characterize the EMCCD, we always separate our results into the 8 outputs.}
   \end{figure}

For all of our conducted tests, we analyze our results for each of the EMCCD's 8 outputs independently.  Figure \ref{fig:outputrep} shows a median bias frame at -45 $^{\circ}$C, which will be the EMCCD's nominal operating temperature. 

\subsection{Readout Noise} \label{ron}
\begin{figure} [ht]
   \begin{center}
   \begin{tabular}{c} 
   \includegraphics[height=7cm]{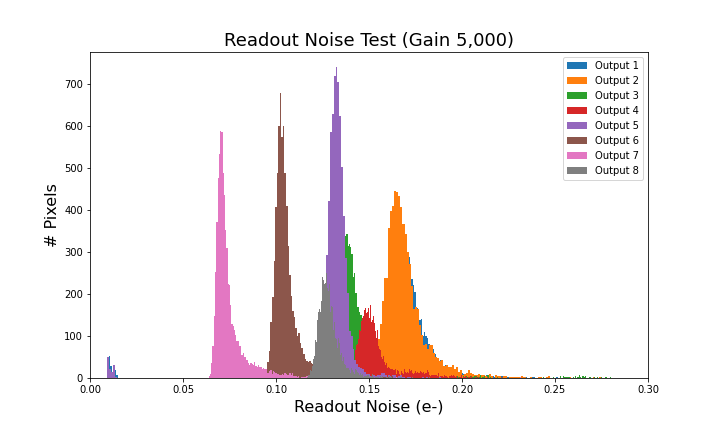}
   \end{tabular}
   \end{center}
   \caption[example] 
   {\label{fig:readoutnoise} 
The readout noise histogram for the 8 outputs of the EMCCD.}
   \end{figure}

We first assess the readout noise of the camera when operated at -45 $^{\circ}$C, at EM gain of 5,000 and FPS of 3,000. The EMCCD must have a low readout noise of $<$ 0.1 e-/pix/frame with this setup in order to detect signals from faint targets. All readout noise tests are performed with dark frames, such that the camera cap is on. We take 1,000 frames, obtaining a histogram of the readout noise. For each output, we subtract the median of dark frames from the 1,000 dark frames and obtain the standard deviation of the 1,000 frames. We then multiply the standard deviation frame by the K-Gain set for each exposure series and divide by the EM gain of 5,000 to obtain units of electrons, and plot a histogram of readout noise for each output. Our results are shown in Figure \ref{fig:readoutnoise}. We also represent the median readout noise for each output in Table \ref{tbl:1}. The readout noise is slightly higher than the expected value given in the datasheet, but below the performance simulations value of 0.4 e-/pix/frame. 

\begin{table}[ht]
\caption{Median Readout Noise of the EMCCD [e-] for each Detector Output} 
\label{tbl:1}
\begin{center}       
\begin{tabular}{|l|l|} 
\hline
\rule[-1ex]{0pt}{3.5ex} Output & Median Readout Noise [e-] \\
\hline
\rule[-1ex]{0pt}{3.5ex} 1 &  0.169787  \\
\hline
\rule[-1ex]{0pt}{3.5ex} 2 &  0.167197  \\
\hline
\rule[-1ex]{0pt}{3.5ex} 3  &   0.139553  \\
\hline
\rule[-1ex]{0pt}{3.5ex} 4  &  0.150889  \\
\hline
\rule[-1ex]{0pt}{3.5ex} 5  &  0.132648  \\
\hline 
\rule[-1ex]{0pt}{3.5ex} 5  &  0.132648  \\
\hline 
\rule[-1ex]{0pt}{3.5ex} 6  &  0.103851 \\
\hline 
\rule[-1ex]{0pt}{3.5ex} 7  &  0.071985  \\
\hline 
\rule[-1ex]{0pt}{3.5ex} 8  & 0.127600 \\
\hline 
\end{tabular}
\end{center}
\end{table}

\subsection{Clock-Induced Charges} \label{cic}

We also test the clock-induced charges (CIC) of the EMCCD. The CIC is a source of noise in the EMCCD where false counts are created when the photoelectrons travel in the EM register (\cite{NuVuCameras}). These charges appear as a false count that cannot be distinguished from the true signal. We perform this test by setting the EMCCD to ``photon counting mode", where we take dark exposures at the max frame rate of 0.33 ms, and bias subtract them. Then, we obtain frames where all of the pixels should have ``0" counts unless a CIC event occurs, in which case the pixel will have a ``1" count.  \par
\begin{figure}  [ht]
   \begin{center}
   \begin{tabular}{c} 
   \includegraphics[width=\columnwidth]{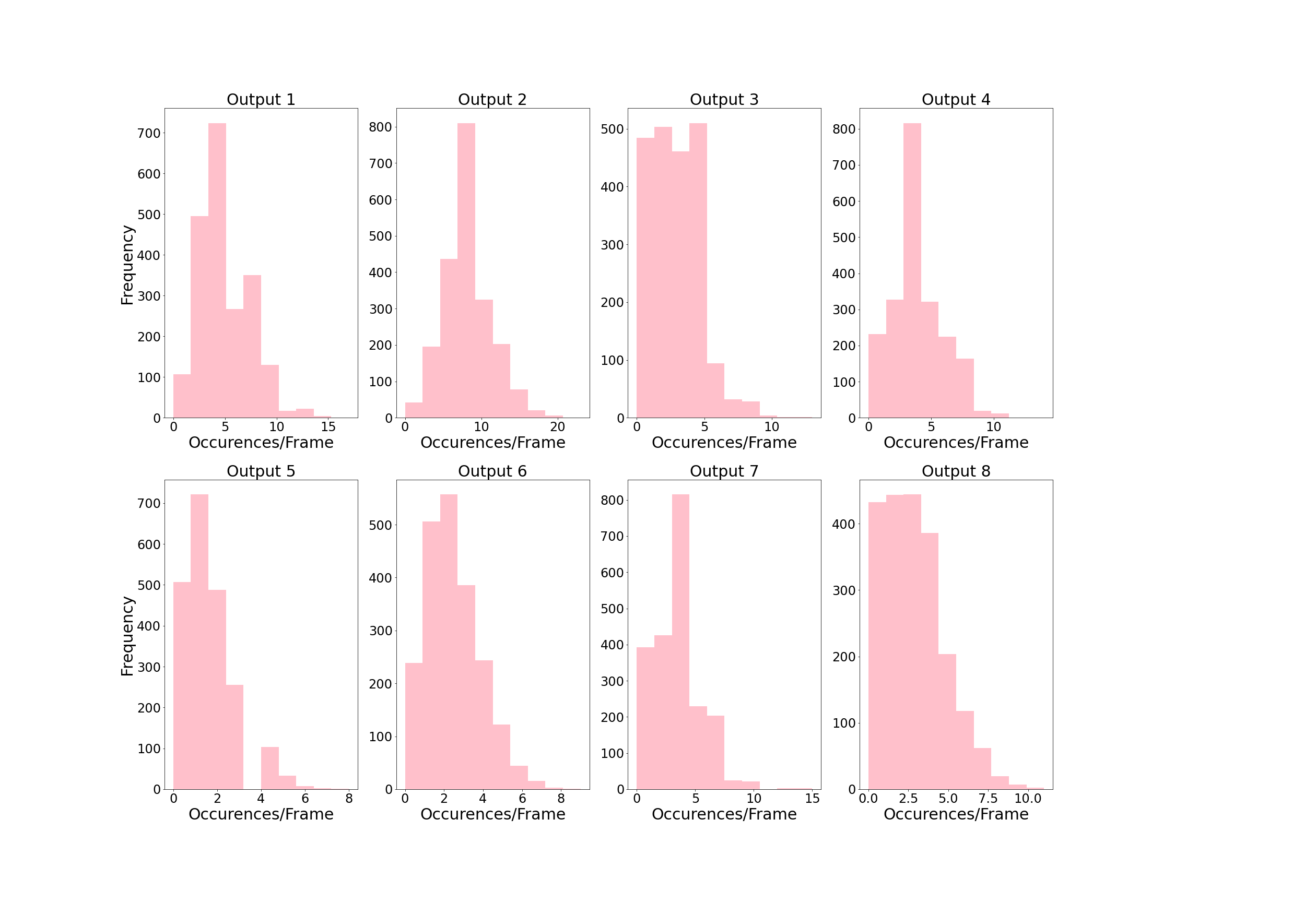}
   \end{tabular}
   \end{center}
   \caption[example] 
   {\label{fig:cic} 
Histogram of the clock-induced charge in the EMCCD for 2,118 total frames at 3,000 FPS and 5,000 EM gain.} 
   \end{figure}

Figure \ref{fig:cic} shows the histogram of these pixels for every output over the 2,118 frames obtained. 

The CIC average for each output is calculated by summing the CIC events for every output over the 2,118 frames and then dividing the sum by the number of frames (2,118) and pixels in each output (120 x 60).
Our results are shown in Table \ref{tbl:2}.

\begin{table}[ht]
\caption{Median CIC for each Detector Output} 
\label{tbl:2}
\begin{center}       
\begin{tabular}{|l|l|} 
\hline
\rule[-1ex]{0pt}{3.5ex} Output & CIC/pixel over 2,118 Frames  \\
\hline
\rule[-1ex]{0pt}{3.5ex} 1 & 0.000695 \\
\hline
\rule[-1ex]{0pt}{3.5ex} 2 &  0.001129  \\
\hline
\rule[-1ex]{0pt}{3.5ex} 3  &   0.000396  \\
\hline
\rule[-1ex]{0pt}{3.5ex} 4  &   0.000536   \\
\hline
\rule[-1ex]{0pt}{3.5ex} 5  &  0.000204 \\
\hline 
\rule[-1ex]{0pt}{3.5ex} 5  &  0.132648  \\
\hline 
\rule[-1ex]{0pt}{3.5ex} 6  &  0.000313 \\
\hline 
\rule[-1ex]{0pt}{3.5ex} 7  &  0.000449 \\
\hline 
\rule[-1ex]{0pt}{3.5ex} 8  & 0.000425 \\
\hline 
\end{tabular}
\end{center}
\end{table}

\subsection{Multiple Regions of Interest} \label{mroi}

   \begin{figure} [ht]
   \begin{center}
   \begin{tabular}{c} 
   \includegraphics[width =\columnwidth]{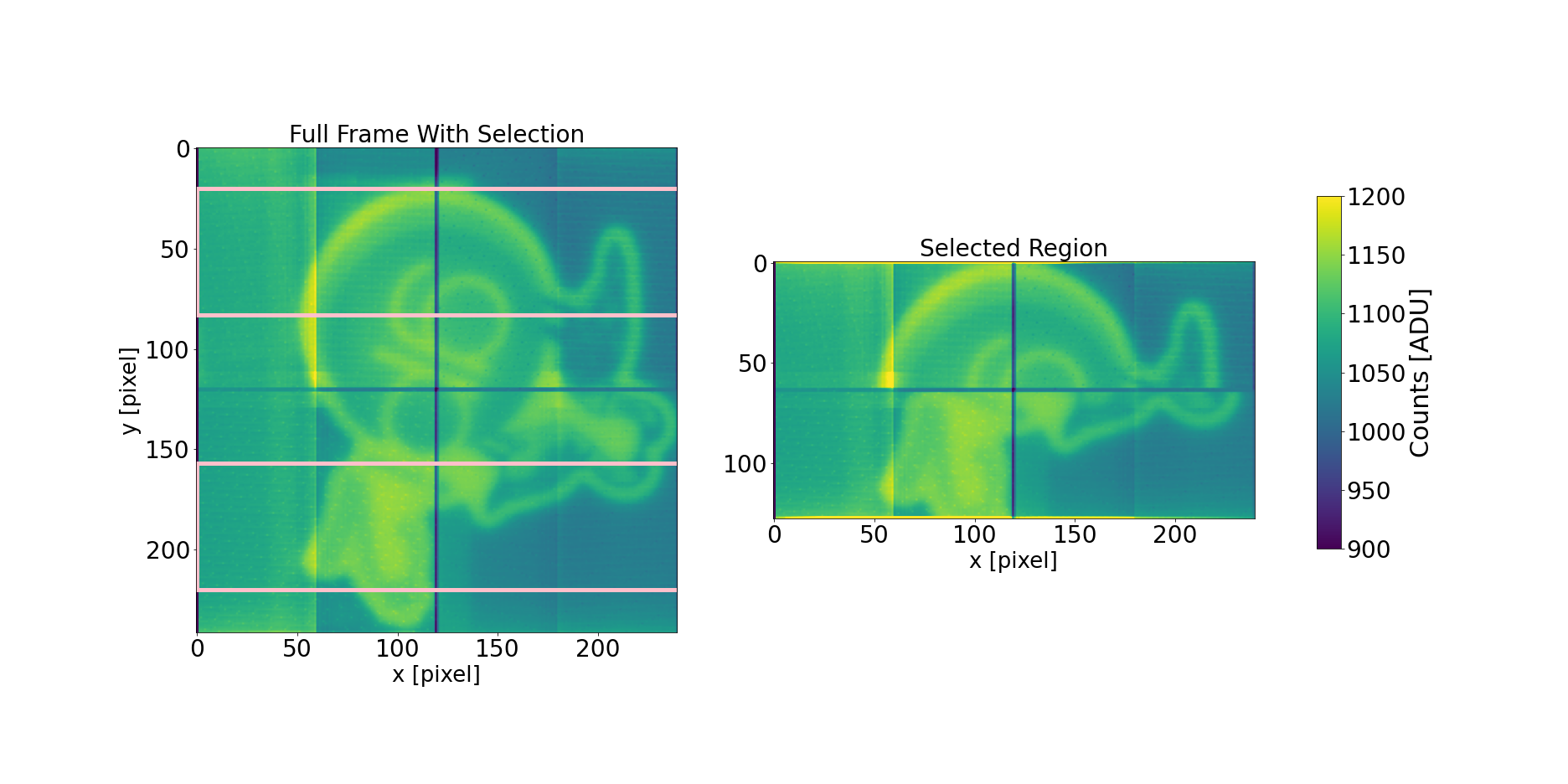}
   \end{tabular}
   \end{center}
   \caption[example] 
   {\label{fig:mroi} 
The multiple regions of interest of our EMCCD. The left figure shows the full frame of the GPI 2 logo with the pink rectangles representing the region selected for mROI. Right figure shows a 63 pixel in height (offset by 20 pixels from the edges) mROI configured by the camera.}
   \end{figure}

We test the multiple regions of interest (mROI) functionality of the EMCCD. The mROI is an important feature for the WFS since the pupils generated by the pyramid are only 63 pixels in height on the detector (offset by 20 pixels from the edges). By selecting the region of interest, one can read out the entire information given by the WFS without having to read the full frame, which allows for a faster frame rate and therefore a better atmospheric correction. We use an image of the GPI 2.0 logo, which is illuminated by an Explore Scientific Astro R-Lite Red Flashlight (Model ES-FL1001), for facilitating the visualization of the feature. We test this by obtaining the full frame and the mROI region and comparing the obtained frame to our own slicing of the full frame for y = 63 (63 pixels vertically) and dy = 20 (20 pixels offset from the top and bottom of the image). The figures match the requested mROI frame. The results are shown in Figure \ref{fig:mroi}. The EMCCD is only capable of performing mROI in the vertical direction, as all pixels in a row must be read. When testing the limitations of the mROI, we note that we can select a region as small as a single row in the EMCCD.

\subsection{Binning} \label{binning}
\begin{figure} [ht]
   \begin{center}
   \begin{tabular}{c} 
   \includegraphics[width = \columnwidth]{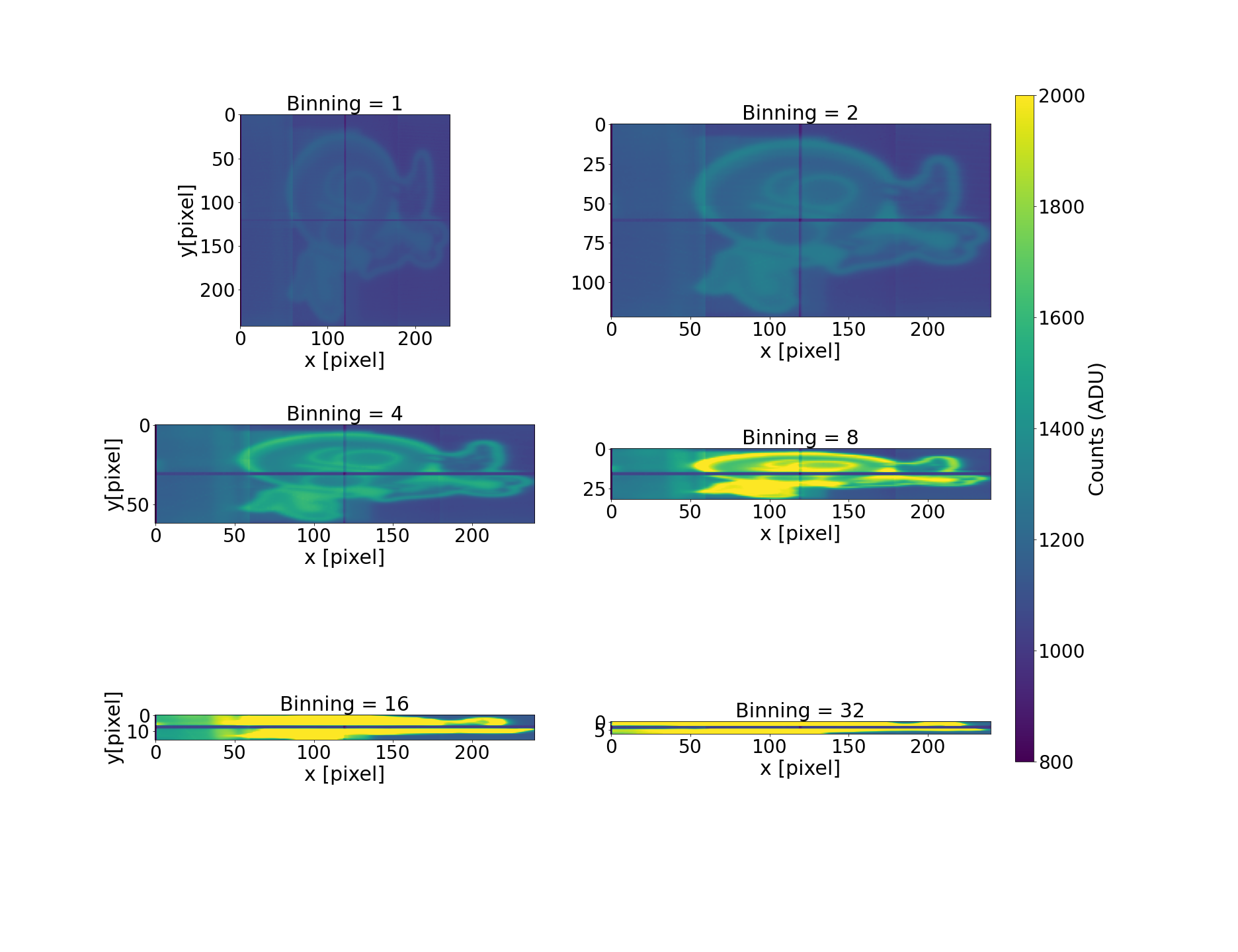}
   \end{tabular}
   \end{center}
   \caption[example] 
   {\label{fig:binning} 
The binning feature of our EMCCD. Binning of the images allows for higher counts in a pixel at the expense of decreased spatial resolution, thus allowing for wavefront correction on fainter targets.}
   \end{figure}

The EMCCD offers binning of 1, 2, 4, 8, 16 and 32x. The binning can only be done in the vertical direction as the pixels in the horizontal direction must be read for the entirety of the row. The test is once again conducted using the GPI 2.0 logo to ease the visualization of the feature. We show the binning feature of the camera in Figure \ref{fig:binning}. The binning feature allows for higher counts in a pixel at the expense of decreased spatial resolution, allowing for wavefront correction on fainter targets.

\subsection{Full Frame Rate} \label{ffr} 
\begin{figure}
    \centering
    \subfloat[\centering ]{{\includegraphics[width=12cm]{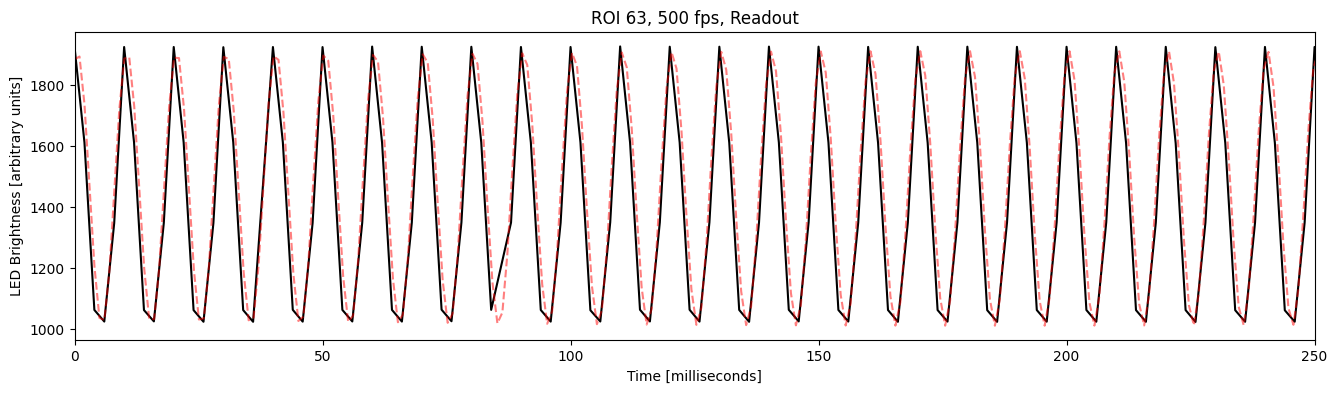} }}%
    \qquad
    \subfloat[\centering ]{{\includegraphics[width=12cm]{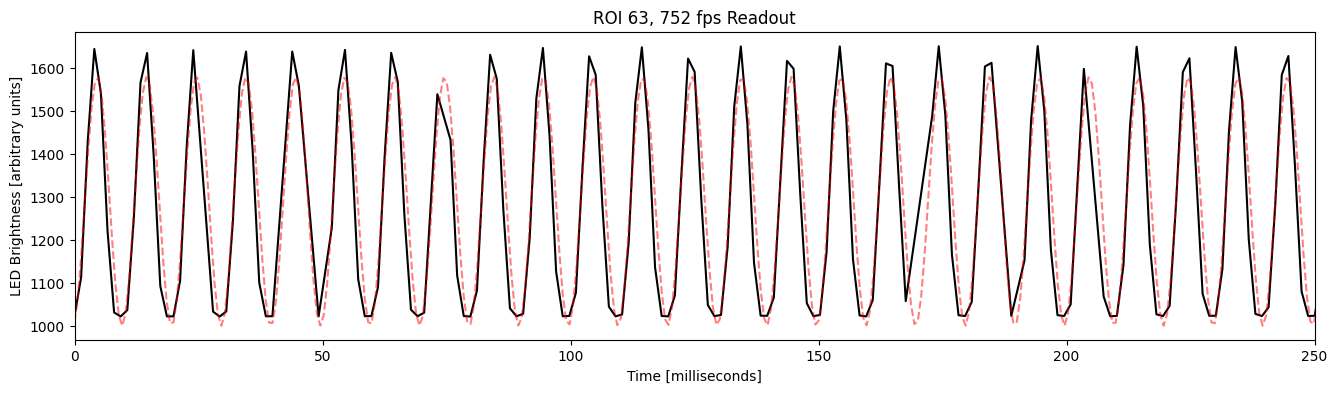} }}%
        \qquad
    \subfloat[\centering ]{{\includegraphics[width=12cm]{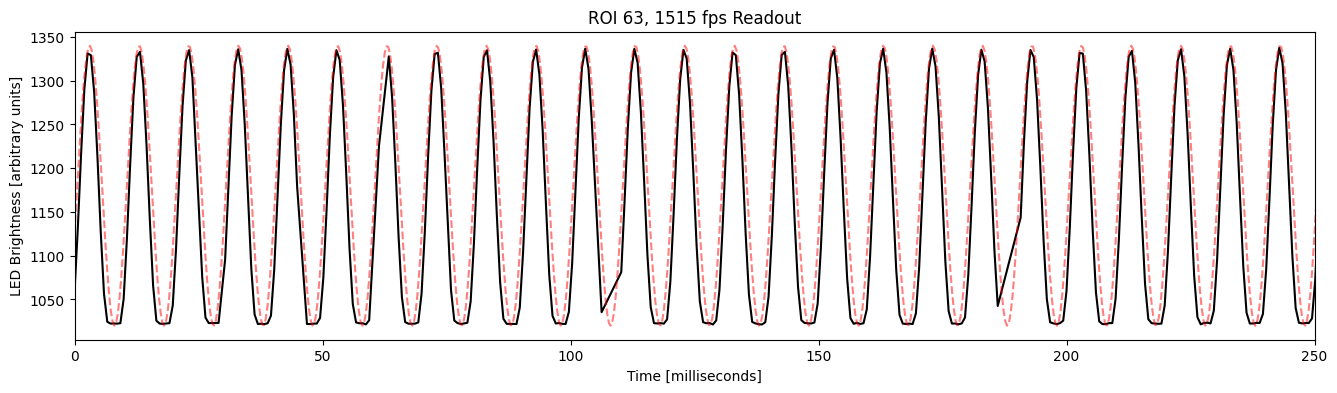} }}%
        \qquad
    \subfloat[\centering ]{{\includegraphics[width=12cm]{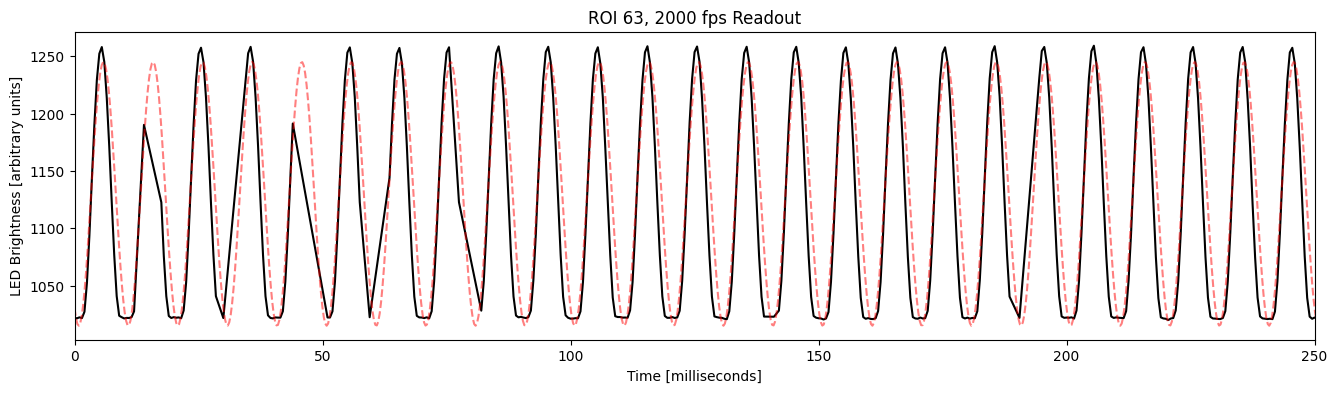} }}%
        \qquad
    \subfloat[\centering ]{{\includegraphics[width=12cm]{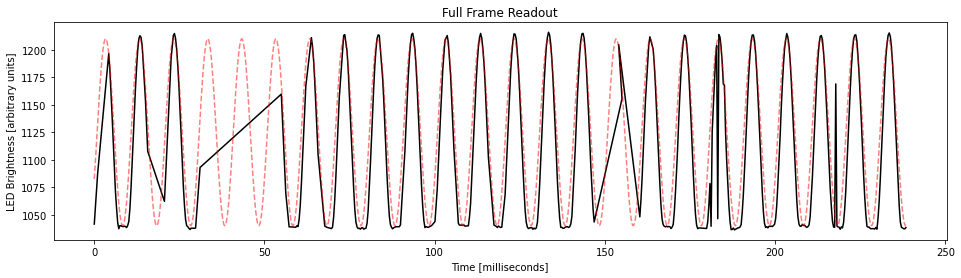} }}%
    \caption{Full frame rate test for the EMCCD using an LED with a sine wave signal. The x-axis shows the time in ms and the y-axis shows LED brightness. The true signal is shown in red while the camera's signal is shown in black. The frame rates are 500 FPS (a), 752 FPS (b), 1515 FPS (c), 2000 FPS (d) and 3000 FPS (e).}%
    \label{fig:ffr}
\end{figure}

We also test for the full frame rate of the EMCCD. This is an important requirement, as it is what determines whether our pyramid wavefront system can operate at the required speed to properly sample atmospheric turbulence. We test this using an LED with a sine wave signal pulsed at a frequency of 100 Hz using an Agilent 33510B waveform generator. We set up an LED with a set sine wave signal that does not saturate the detector, with the sine wave signal slow enough such that the camera FPS is sampling well above the Nyquist frequency. The data is then cropped to only show the image of the LED. We average the values within the cropped image and plot the average brightness given the time of the image, using the time from the header of the images. We perform this test for five different frame rate values: 500, 752, 1515, 2000 and 3000 FPS. We obtain frames using the camera's provided SDK. \par

Our results are shown in Figure \ref{fig:ffr}. We note that 3,000 FPS was never fully reached, which we analyze by comparing the true input signal to the signal obtained from the camera data. As the FPS increases, we get incrementally worse latency between frames. Note the jagged edges present in every time series plot, indicating that we are not sampling the data consistently. While it appears that the camera is able to achieve the speed that we need, there are times when frames are dropped and the timing between frames is not consistent. \par

\subsection{Flat Field Tests} \label{flats}
We conduct flat fielding tests for our EMCCD. For this, we utilize the Newport 819D-SL-3.3 integrating sphere, used in
conjunction with a Newport 6332 quartz tungsten halogen lamp, a white light lamp operated at 50 W and a Newport 60043 socket
adaptor. The bulb allows for the change in light intensity using a Kikusui Stabilized power supply (Model PAB 8-2.5). We test the EM gain, exposure time and light level linearities for the EMCCD. For all of our tests, we subtract the bias frame with matching EM gain and use the median of a cube with 300 frames. 

\subsubsection{EM Gain Linearity} \label{emgainlinear}
\begin{figure}
    \centering
    \subfloat[\centering ]{{\includegraphics[width=8.1cm]{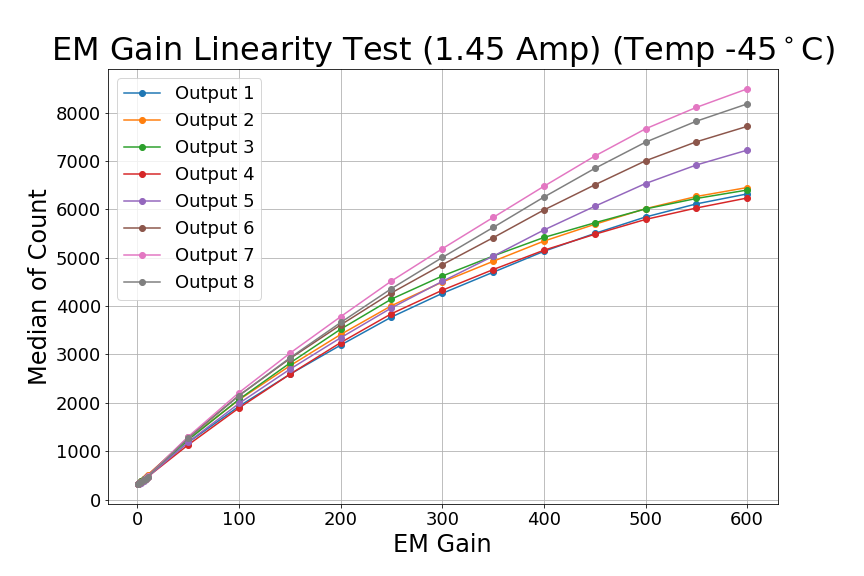} }}%
    \qquad
    \subfloat[\centering ]{{\includegraphics[width=8.1cm]{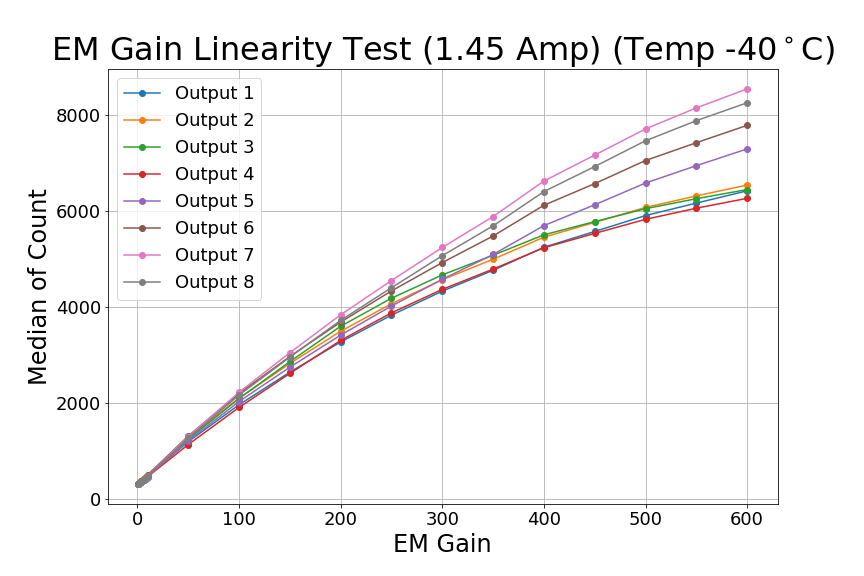} }}%
        \qquad
    \subfloat[\centering ]{{\includegraphics[width=8.1cm]{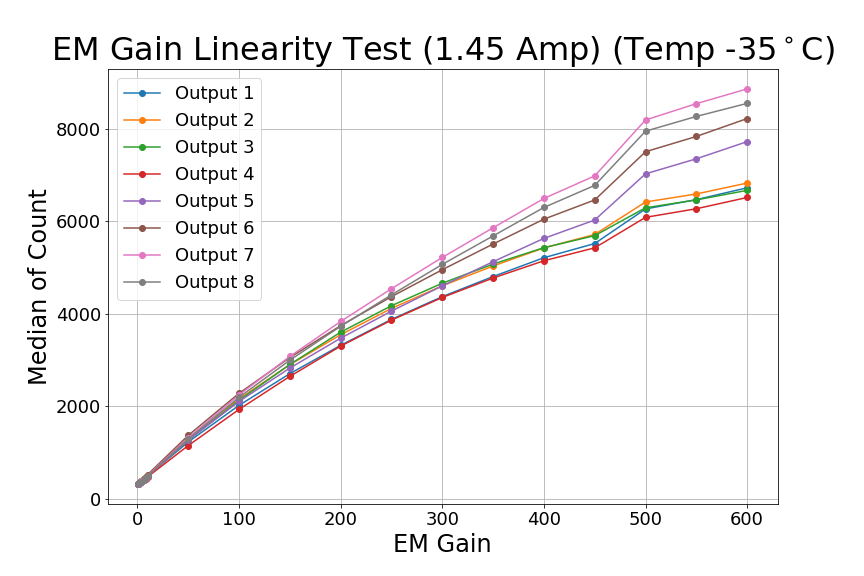} }}%
        \qquad
    \subfloat[\centering ]{{\includegraphics[width=8.1cm]{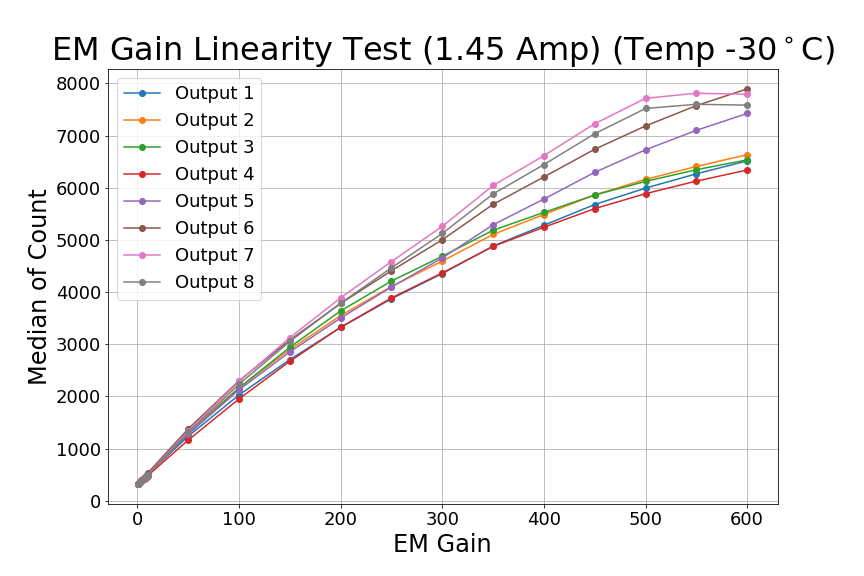} }}%
        \qquad
    \subfloat[\centering ]{{\includegraphics[width=8.1cm]{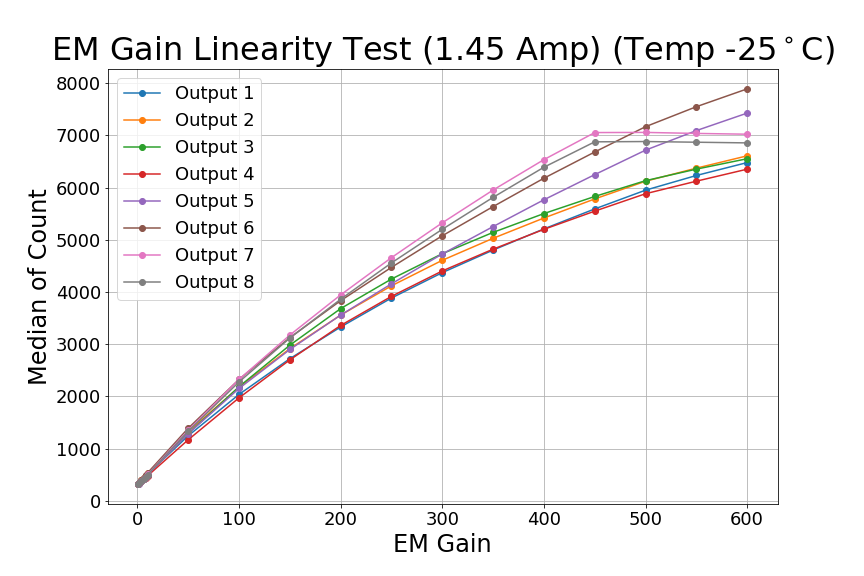} }}%
    \caption{EM Gain Linearity test with matching light levels but varying temperatures for the EMCCD. The median of counts is shown as a function of EM gain. The temperatures are -45$^{\circ}$C (a), -40$^{\circ}$C (b), -35$^{\circ}$C (c), -30$^{\circ}$C (d) and -25$^{\circ}$C (e).}%
    \label{fig:emlinear}
\end{figure}

We first test the linearity with changing EM gain for the camera at higher light levels (1.45 Amps). We perform these tests to characterize the camera's behavior as a function of change in the EM gain, since different target magnitudes will require different EM gain configurations. We test this for chip temperatures of -45, -40, -35, -30 and -25 $^{\circ}$C for EM gain 1 to 600. We do not go further than that so that we do not saturate the outputs in the EMCCD, as that can damage the detector. We plan on operating the PWFS at -45 $^{\circ}$C; however, this test allows us to characterize the camera's dependence on temperature. We subtract the bias with matching EM gain from each cube with 300 frames, then obtain the median of each output.

\begin{figure} [ht]
   \begin{center}
   \begin{tabular}{c} 
   \includegraphics[height=7cm]{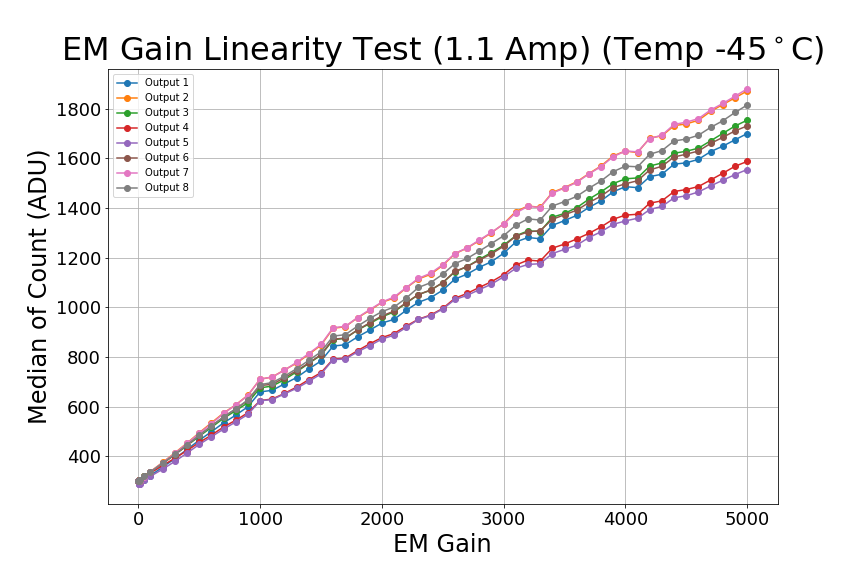}
   \end{tabular}
   \end{center}
   \caption[example] 
   {\label{fig:emgain1amp} 
 EM Gain Linearity test with lower light levels (non-saturating) for the EMCCD. The median of counts is shown as a function of EM gain, in ADU. Here, we go to the full EM gain range of 1 - 5,000. We find a linear behavior with no saturation of the detector.}
   \end{figure}

We find that Outputs 7 and 8 reach saturation at higher temperatures (mainly -25 and -30 $^{\circ}$C). 
We then test the camera's EM gain linearity at a lower light level (1.1 Amp) for the operating temperature of -45 $^{\circ}$C, and find that at lower light levels saturation is no longer reached and that the EM gain behavior is linear. The results are shown in Figure \ref{fig:emgain1amp}.

 \subsubsection{Exposure Time and Light Level Linearity} \label{timelinearity}

 \begin{figure} [ht]
    \centering
    \subfloat[\centering ]{{\includegraphics[width= 8cm]{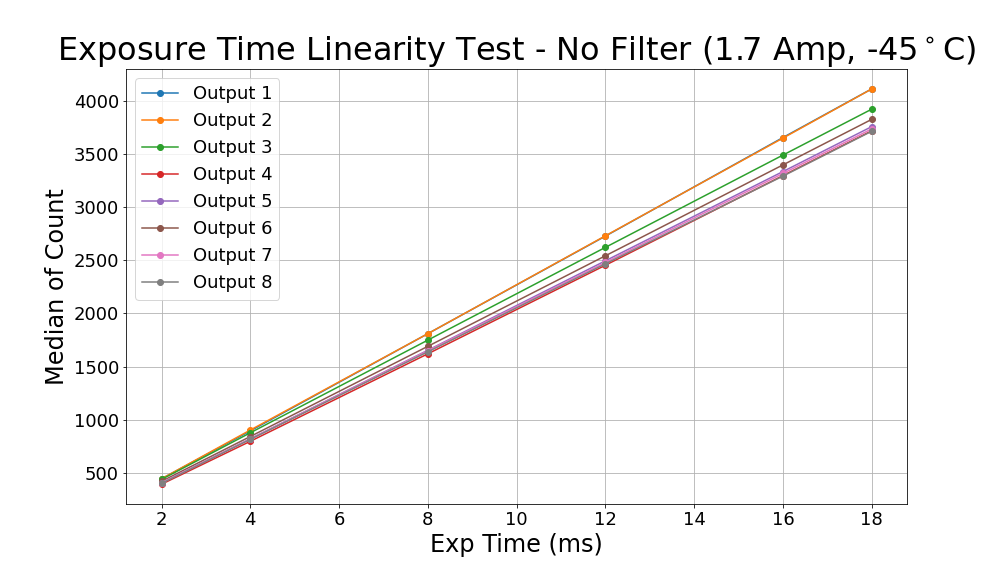} }}%
    \qquad
    \subfloat[\centering ]{{\includegraphics[width=8cm]{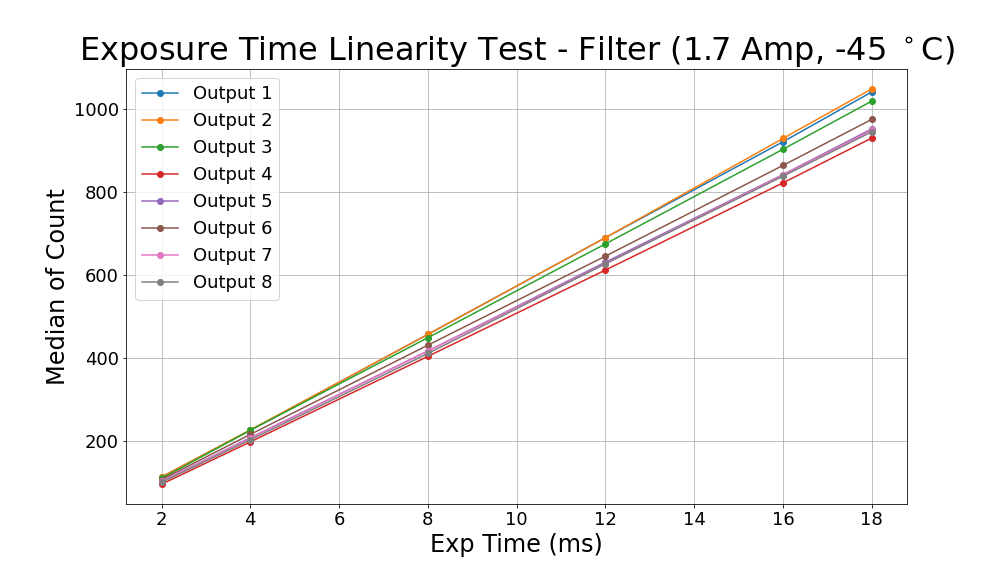} }}%
    \caption{The median counts in ADU for varying exposure time in our EMCCD. The tests were conducted for light levels without an ND filter (a) with an ND Filter (b). A linear behavior is achieved for both cases. }%
    \label{fig:exptime}
\end{figure}

We test the exposure time linearity of our EMCCD for two light levels: by setting our flat lamp to 1.7 Amps and placing a Thorlabs Neutral Density (ND) 0.6 filter in front of the camera. This should decrease light levels by a factor of $\sim$ 4. Our exposure times used are 2, 4, 8, 12, 16 and 18 ms. We characterize this feature to ensure we have an expected behavior as we change the camera's frame rate. 
Results are presented in Figure \ref{fig:exptime}. We find that for both light levels the behavior of all outputs is linear with changing exposure time.

\subsubsection{Light Level Ratios} \label{ratios}
\begin{figure} [h!]
   \begin{center}
   \begin{tabular}{c} 
   \includegraphics[height=7cm]{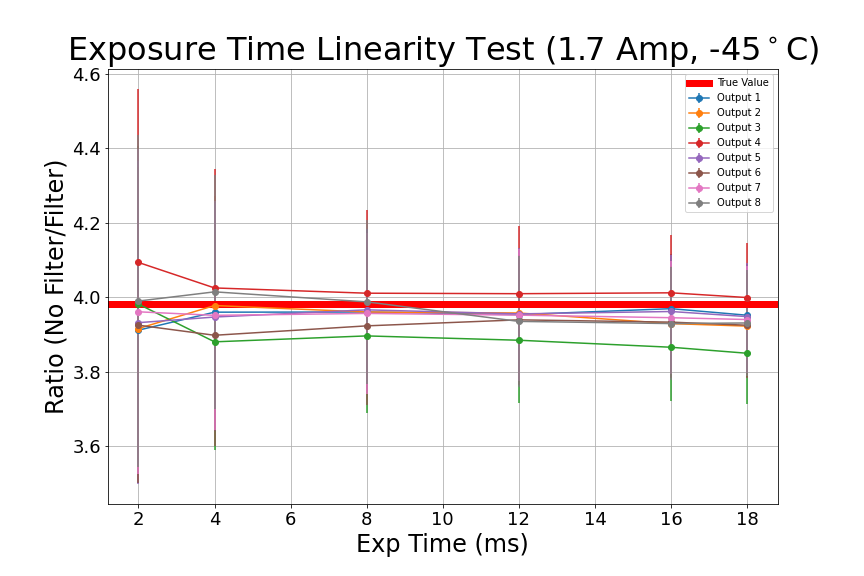}
   \end{tabular}
   \end{center}
   \caption[example] 
   {\label{fig:ratiolevel} 
The ratio of light levels for flat fields with and without an ND filter. The true value is plotted as the red horizontal line, while measured levels are shown as scatter points with error bars for each output. Error bars are calculated using the Poisson noise for each output.}. 
   \end{figure}

We test that the counts obtained by the EMCCD correspond to the expected ratio given by our ND 0.6 filter (which corresponds to a decrease in light levels by a factor of 3.981). In order to do that, we plot the ratio of counts given by the two individual light levels at corresponding exposure times. Our results presented in Figure~\ref{fig:ratiolevel} illustrated that the ratio of the two light levels gives the expected filter value.

\section{CONCLUSION \& FUTURE WORK} \label{conclusion}
In this work, we evaluate the HNü 240 EMCCD's performance. Specifically, we verify that the detector's readout noise, clock-induced charges, EM gain linearity and exposure time linearity meet the requirements for proper integration with the GPI 2.0 pyramid wavefront sensor system. We find that the camera has the expected readout noise, clock-induced charges, and EM gain/exposure time linearity behaviour. However, we do not obtain the expected full frame rate results for the EMCCD, likely because we were using the SDK provided to obtain the frames for the test shown in Section \ref{ffr}. Our findings and requirements are summarized in Table \ref{tbl:3}. In order to reach the full frame rate required for GPI 2.0, the camera's serial commands paired with the Pleora SDK must be employed. We will present the full frame rate results with serial commands in future work. \par

\begin{table}
\caption{Table with requirement tests for the EMCCD and their results.} 
   \begin{threeparttable}
   \begin{center}
\label{tbl:3}

\begin{tabular}{@{} p{4cm}|p{4cm}|p{4cm}|p{1cm} @{}}

 \hline
 Test & Requirement & Results & Pass or Fail? \\ [0.5ex] 
 \hline\hline
Readout Noise &  $<$ 0.1 e- readout noise at 3,000 FPS and EM gain of 5,000 & Median of 0.127600 e- & Pass  \\
\hline
Binning  & Binning Capability of 1, 2, 4, 8, 16, 32x & Binning Capability was achieved for all values & Pass \\
\hline
mROI  &  Configure the mROI for the 4 pupils of the EMCCD & mROI region was achieved with correct pixel offset and location & Pass  \\
\hline
Full Frame Rate & Full Frame Rate of 3 kHz & Full Frame Rate not achieved with SDK/GUI & $Fail^{*}$ \\
\hline
EM Gain Linearity  & EM gain functionality from 1 - 5,000 & Linearity Found for EM gain up to 5,000 when detector is not saturated & Pass \\
\hline
Exposure Time and Light Level Linearity  & The behavior of counts for exposure time and light levels must be linear & Linearity is achieved for all outputs & Pass \\
\\
 \hline

\end{tabular}
\begin{tablenotes}
   \item[*] Pending serial command test to achieve full frame rate.  
  \end{tablenotes}
\end{center}
   \end{threeparttable}
\end{table}

All of the tests conducted in this work were performed before the alignment and integration of the pyramid wavefront sensor. Therefore, in a future work, we hope to present on the performance of the pyramid wavefront sensor system alignment and the EMCCD's characterization after the system's integration (e.g. the full frame rate test and characterization of the pyramid's pupils).

\acknowledgments 
GPI 2.0 is funded by in part by the Heising-Simons Foundation through grant 2019-1582. The GPI project has been supported by Gemini Observatory, which is operated by AURA, Inc., under a cooperative agreement with the NSF on behalf of the Gemini partnership: the NSF (USA), the National Research Council (Canada), CONICYT (Chile), the Australian Research Council (Australia), MCTI (Brazil) and MINCYT (Argentina). Portions of this work were performed under the auspices of the U.S. Department of Energy by Lawrence Livermore National Laboratory under Contract DE-AC52-07NA27344.

\clearpage
\printbibliography

\end{document}